%% file: arXiv.tex
\documentclass[twocolumn]{article}
\usepackage[ top=2cm, bottom=2cm, left = 1.5cm, right = 1.5cm]{geometry}
\input{astro_journal_symbols}

\usepackage{multicol}
\usepackage{abstract} 
\usepackage{float}
\usepackage{caption}
\usepackage{graphicx}	

\usepackage{natbib}   
\usepackage{physics}    

\usepackage[T1]{fontenc} 
\usepackage{txfonts} 

\usepackage[switch,mathlines]{lineno}

\usepackage[bottom]{footmisc} 
\usepackage{hyperref}

\makeatletter
\@ifpackageloaded{hyperref}{}{\newcommand{\href}[1]{}}
\makeatother

\newcommand{\alttext}[1]{}

\newcommand{\ly}{Ly$\alpha$ }

\usepackage{titlesec}
\titleformat{\section}{\large\bfseries}{\thesection}{1em}{}
\titleformat{\subsection}
  {\normalfont\large}  
  {\thesubsection}{1em}{}
\usepackage[bottom]{footmisc}  

\usepackage{authblk} 
     
\title{ Impact of Ly $\alpha$ radiation force on super-Eddington accretion onto a massive black hole}
\author[1]{\textbf{Takuya \textsc{Mushano}}}
\author[1,*]{\textbf{Takumi \textsc{Ogawa}}}
\author[1]{\textbf{Ken \textsc{Ohsuga}}}
\author[1]{\textbf{Hidenobu \textsc{Yajima}}}
\author[2]{\textbf{Kazuyuki \textsc{Omukai}}}

\affil[1]{Center for Computational Sciences, University of Tsukuba, Ten-nodai, Tsukuba, Japan}
\affil[2]{Astronomical Institute, Graduate School of Science, Tohoku University, Sendai, Japan}
\affil[*]{{E-mail: takumi@ccs.tsukuba.ac.jp}}

\begin{document}
\twocolumn[
\begin{@twocolumnfalse}
\maketitle
\begin{abstract}
        \noindent\textbf{\large Abstract}        \vspace{1em} \\  
{\normalsize
\input{abstract}
}
\end{abstract}   
\vspace{1em} 
    \vspace{1em}
\end{@twocolumnfalse}
]

\section{Introduction}
\input{intro_E}

\section{Basic Equations and Numerical Methods}
\input{method_E}

\section{Results}
\input{result_E}

\section{Summary and Discussion}
\input{summary_and_discussions}

\section*{Acknowledgement}
The authors wish to express their cordial gratitude to their most revered mentor, Prof. Masayuki Umemura, the {\it Pater Patriae} of the Theoretical Astrophysics Group at the University of Tsukuba, for his continual interest, advice, and encouragement.
This work is supported by JSPS KAKENHI Grant Numbers JP22KJ0420 (TM), in part by MEXT/JSPS KAKENHI grant numbers 17H04827, 20H04724, 21H04489 (HY), 	17H01102, 22H00149 (KO),
NAOJ ALMA Scientific Research grant numbers 2019–11A, JST FOREST Program, grant number JP-MJFR202Z, and Astro Biology Center Project research AB041008 (HY).


\bibliographystyle{mnras}
\bibliography{library,mybook,references} 

\appendix
\input{appendix_velocitytest.tex}

\input{appendix_compare_homogeneous}

\end{document}

%% file: astro_journal_symbols.tex
\newcommand{\aj}{AJ}
\newcommand{\araa}{ARA\&A}
\newcommand{\apj}{ApJ}
\newcommand{\apjl}{ApJL}

\newcommand{\mnras}{MNRAS}

\newcommand{\pasj}{PASJ}

\newcommand{\ssr}{Space\ Sci.\ Rev.}

\newcommand{\nat}{Nature}

%% file: abstract.tex
The viability of super-Eddington accretion remains a topic of intense debate, crucial for understanding the formation of supermassive black holes in the early universe.
However,
the impact of \ly radiation force on this issue remains poorly understood.
We investigate the propagation of the \ly photons 
and evaluate the \ly radiation force within a spherically symmetric accreting HI gas onto the central black hole. We solve the radiation transfer equation, incorporating the destruction processes of \ly photons through two-photon decay and collisional de-excitation.
We find that the \ly photons, originating in the HII region around black holes, suffer from multiple resonance scattering before being destroyed via two-photon decay and collisional de-excitation. 
Hence, the \ly radiation force undergoes a significant amplification, surpassing gravity at the innermost section of the HI region.
This amplification, quantified as the force multiplier, reaches approximately 130 and remains nearly constant, regardless of the optical depth at the line center, provided the optical thickness of the flow is within the range of $10^{10-14}$.
The requisite lower limit of the product of gas density and black hole mass to realize the super-Eddington accretion is found to be in the range $(2-{40}) \times 10^9 M_\odot\,{\rm cm}^{-3}$, which is a few to tens of times larger than the minimum value obtained without accounting for the \ly radiation force.
The pronounced amplification of the \ly radiation force poses a substantial challenge to the feasibility of super-Eddington accretion. 

%% file: intro_E.tex
{
Supermassive black holes (hereafter, SMBHs) ubiquitously {exist} at the center{s} of large galaxies. 
The tight correlation of {their} masses with those of host galaxies strongly suggests their coevolution \citep[][for the review]{Kormendy2013}. 
Despite decades of intensive investigations into the origin of those SMBHs, it still remains as one of the most significant enigmas in astrophysics. The discoveries of SMBHs in the early universe \citep{Fan2004,Mortlock2011,Wu2015,Matsuoka2018a,Bogdan_et_al2023} provide very strong constraints on the timescale of their growth. 
While various hypotheses exist regarding the formation of SMBHs \citep[][for the review]{Volonteri2012,Haiman2013,Inayoshireview2020}, {
the super-Eddington growth of seed black holes (BHs) {in massive, dense atomic gas clouds }emerges as one of the most promising scenarios.}
}

Super-Eddington accretion flow has been investigated by a number of authors. Within the scale extending from the event horizon to the centrifugal radius, an accretion disk forms around a BH. At this scale, super-Eddington accretion occurs naturally in environments characterized by abundant gas supply, driven by the interplay of photon trapping and the anisotropy of radiation 
(\citealt{Watarai2000,Ohsuga_et_al.2005,McKinney_et_al.2014,Jiang_et_al.2014,Sadowski_et_al.2015}; 
and \citealt{Ohsuga_and_Mineshige.2014} for the concise review). 

On larger scales, comparable to the Bondi radius, the situation differs significantly. {
The radiation originating from the vicinity of the central black hole is expected to be approximately spherically symmetric, primarily due to electron scattering, ionization, and recombination processes. Although some anisotropy may exist around the black hole, the mechanical forces are also spherically symmetric, as the centrifugal force becomes nearly negligible in regions far from the centrifugal {radius}.
 Therefore, i}nstead of disk accretion, a roughly spherical accretion flow is realized, leading to a more pronounced influence of radiation pressure and heating on the accretion dynamics. 
A number of investigations by different authors have confirmed that accretion rates in such flows are significantly lower than the Eddington rate 
(\citealt{Milosavljevic2009a,Milosavljevic2009b,Park2011,Park2012,Park2013}). Later, \cite{inayoshi2016hyper}{, performing one-dimensional spherically symmetric radiation hydrodynamic simulations
incorporating photo-ionization from the Bondi {scale down to the photon trapping radius},} reported that the accretion rate becomes comparable to the Bondi rate, which as large as thousands of times the Eddington rate, provided the condition $n_{\infty}M_{{\rm BH}}> 10^{9} \:M_\odot{\rm cm^{-3}}$ is satisfied. Here $n_{\infty}$ is the gas density at the infinity and $M_{{\rm BH}}$ is the BH mass.
This condition is determined {by {the comparison between the sizes of the Str\"{o}mgren and Bondi radii}}. 

{Their results are particularly noteworthy as they reveal the possibility of super-Eddington accretion even under the most conservative assumption of spherical accretion.}
 Several authors also investigated the condition under which super-Eddington accretion is realized in this context \citep{sakurai2016hyper,Yajima2017,Takeo2018,Sugimura2017,Toyouchi2021}. {They report that radiation pressure is rather subdominant and radiation mainly affects dynamics via {photo-ionization} heating of the inflowing gas, by taking into account the electron scattering and the bound-free absorption of the primordial gas in estimating the radiation force.}  However, the pressure of {Ly$\alpha$ photons}, generated through the 
 recombination of hydrogen atoms, may impact gas dynamics significantly as it is amplified by multiple resonance scatterings \citep{Milosavljevic2009a,smith2015lyman}. The Ly$\alpha$ radiation force increases with the increase of the typical optical depth of a system (\citealt{Adams1975,smith2016evidence,dijkstra2006lyalpha} and \citealt{Dijkstra_review2019} for the review). For example, in a massive halo with the line-center optical depth of $10^{11}$ and a temperature of $10^4{\:\rm K}$, the force can be amplified by $10^3$ times compared to the single-scattering case.

The destruction process of Ly$\alpha$ photons, such as the two photon decay and the collisional de-excitation, can be important in the context of the super-Eddington growth of massive seeds. 
Although the two-photon decay is a forbidden transition from the 2s to 1s levels of atomic hydrogen and its probability is much smaller than that of resonant scattering, it becomes significant in highly optically thick media.
This is because the Ly$\alpha$ photon undergoes numerous scatterings before escaping to the outside.
The decrease in the number of Ly$\alpha$ photons due to collisional de-excitation is also non-negligible in dense media.

In this paper we investigate the effect of Ly$\alpha$ radiation on {the Bondi accretion flows{{} onto intermediate-mass black holes} under fixed gas properties, such as density, velocity and temperature.} {We solve the Ly$\alpha$ radiative transfer by employing the radiation diffusion and Fokker-Planck approximations \citep{neufeld1990transfer}}. We consider the destruction mechanisms of the Ly$\alpha$ photons by the two-photon decay and the collisional de-excitation, and compare the radiation force to the gravity to estimate whether super-Eddington accretion is feasible or not.

This paper is organized as follows: In Section 2, we introduce the basic equations and assumptions of our models. Section 3 presents our results and discusses the feasibility of super-Eddington accretion, and finally in Section 4, we summarize our study. 

%% file: method_E.tex
{We suppose a 
spherical accretion flow of pure hydrogen gas onto a BH at the center. Surrounding the BH, there should be a spherical ${\rm HII}$ region{{}, extending sufficiently far from the BH,} created by the ionizing photons emitted from the circum-BH accretion disk. Enormous Ly$\alpha$ photons are emitted outward from the ${\rm HII}$ region and then undergo multiple scattering by purely atomic hydrogen gas. 
Hence we set 
the innermost region of our simulation domain to correspond to the outermost region of the ${\rm HII}$ region.

We assume that the flow within the domain follows a spherically symmetric {isothermal }Bondi accretion pattern{{} with the typical temperature of atomic cooling halos ($10^4{\rm K}$), emulating the super-Eddignton flow demonstrated in \cite{inayoshi2016hyper}}. By fixing the distribution of fluid variables, such as fluid velocity, density, and temperature, based on the super-Eddington isothermal Bondi accretion of purely atomic hydrogen gas onto a central black hole, we investigate the steady distributions of Ly$\alpha$ photons. Additionally, we assess the radiation force resulting from the resonance scattering of Ly$\alpha$ photons.
}
Given that the Ly$\alpha$ radiation from the innermost region (i.e. the ${\rm HII}$ region) originates from the BH accretion disk,
the Ly$\alpha$ luminosity should depend on the accretion rate. 
The details of our calculation setup are described below.
\subsection{Basic equations}

In order to account for the multiple scattering of Ly$\alpha$ photons in highly optically thick flows,
we solve {the frequency-dependent $0$-th moment equation of radiation, 
in which the radiation diffusion approximation and the Fokker–Planck approximation are employed, with terms up to $O(\varv/c)$ being taken into account.} 
Since the photon diffusion time scale 
is much shorter than the dynamical time
 in the present situation, the radiation fields can be regarded to become immediately steady. 
{Applying these approximations and assumptions to Eq.(95.18) (95.19) of \cite{Mihalas1984}, the basic equation to be solved here is}
   \begin{align}
      &-\frac{1}{r^2}\pdv{r}\qty(\frac{c{r^2}}{3\kappa_{x_0}} \pdv{E_{x_0}}{r})
        \notag\\
      &+\frac{1}{r^2}\pdv{r}\qty(r^2\varv E_{x_0})+\frac{1}{r^2}\pdv{r}\qty(r^2\varv)\frac{1}{3}E_{x_0} \notag\\
      &-\frac{1}{r^2}\pdv{r}\qty(r^2\varv)\pdv{x_0}\qty[\qty(\frac{c}{\varv_{\rm th}}+x_0)\frac{1}{3}E_{x_0}]\notag\\
      &=
           -\kappa_{x_0}^{\rm a} cE_{x_0} + 4\pi\eta_{x_0}^{\rm r} +\frac{1}{2}\pdv{x_0}\qty(c\kappa^{\rm s}_{x_0}\pdv{E_{x_0}}{x_0}),
      \label{eq:BasicEq}
  \end{align}
where 
{
 $r$,
$c,\varv$ and $\varv_{\rm th}$ are 
the distance from the origin (BH), 
the speed of light, the fluid velocity and the thermal velocity in the gas, respectively. The normalized frequency $x_0$ is defined as $x_0 = (\nu_0-\nu_\alpha)/\Delta \nu_D$, where $\nu_0, \nu_\alpha$ and $\Delta \nu_D$ are the frequency of photons in the fluid comoving frame, }the frequency at the Ly$\alpha$ line center, $\nu_\alpha = 2.466 \times 10^{15} {\rm Hz}$, and the typical Doppler shift by the thermal motion of gas around the Ly$\alpha$ line, {$\Delta \nu_D = \varv_{\rm th}\nu_\alpha/c$}, respectively. 
{
$\kappa^a_{x_0},\kappa^s_{x_0},\kappa_{x_0}$ and $\eta^r_{x_0}$ are the absorption coefficient related to the two-photon decay and the collisional de-excitation, the scattering coefficient of the resonance scattering, {the total opacity defined by $\kappa_{x_0}=\kappa_{x_0}^s+\kappa_{x_0}^a$} and the emission coefficient, respectively, all of which are measured in the fluid comoving frame and at normalized frequency of $x_0$.
}
 The radiation energy density $E_{x_0}$ is defined in the comoving frame as $E_{x_0}\equiv \int \frac{1}{c} I_{x_0} d\Omega_0$ where $I_{x_0}$ is the intensity in the comoving frame at the frequency $x_0$, where $\Omega_0$ is the solid angle measured in the comoving frame. Eq.\ref{eq:BasicEq} includes effects related to the fluid velocity, such as trapping effects by the inflowing medium, the work from/to fluid and the Doppler effects by the velocity dispersion, which correspond to the second, third and fourth terms of the left-hand side, respectively.

The scattering coefficient {for the pure resonant scattering, $\bar\kappa^s_{x_0}$}, is defined as ${\bar\kappa_{x_0}^s}\equiv \sigma_0 H(a,x_0) n_{\rm HI}$ where $\sigma_0,H(a,x_0)$ and $n_{\rm HI}$ are the cross-section at the Ly$\alpha$ line center, the Voigt function, and the number density of the neutral hydrogen gas, respectively. The cross-section $\sigma_0$ and the Voigt function are defined by
 \begin{equation}
    \sigma_{0}=f_{12} \frac{\sqrt{\pi} e^{2}}{m_{e} c \Delta \nu_{D}} \approx 5.88 \times 10^{-14} \qty(\frac{T}{10^4 {\rm K}})^{-1/2}{\rm~cm^2},
  \end{equation}
  \begin{equation}
    H(a,x)=\frac{a}{\pi} \int_{-\infty}^{\infty} \frac{e^{-y^{2}}}{(x-y)^{2}+a^{2}} dy,
  \end{equation}
where $e,m_e,f_{12}$ and $a$ are the elementary charge, the electron mass, the oscillator strength, $f_{12}=0.4162$, and the Voigt parameter, $a = \Delta \nu_{\rm L}/(2\Delta \nu_D)$, which is the ratio between the natural width {$\Delta \nu_{\rm L}$} and the thermal width of the Ly$\alpha$ line, respectively.

The absorption coefficient $\kappa_{x_0}^a$, due to the two-photon decay and the collisional de-excitation, is written as $\kappa^a_{x_0}=p_{\rm dest} \bar\kappa_{x_0}^s$ where the photon destruction probability $p_{\rm dest}$ is the probability 
 that a Ly$\alpha$ photon is annihilated 
at a single resonance-scattering. {With photon destruction, the scattering coefficient $\kappa^s$ becomes slightly different from that for pure resonant scattering $\bar\kappa^s$ and can be written as $\kappa^s_{x_0}=(1-p_{\rm dest})\bar\kappa^s_{x_0}$.}
{Assuming that the hydrogen is in the ground state or in the first excited state and that the higher excited states can be ignored, the photon destruction probability} 
can be calculated as 
\begin{equation}
    p_{\rm dest}=\frac{A_{2\rm ph}+C_{21}}{A_{\rm Ly\alpha}+A_{2\rm ph}+C_{21}}.
\label{eq:p_dest}
\end{equation}
where $A_{{\rm Ly}\alpha}$ is the rate of {the ${\rm Ly}\alpha$ production by spontaneous emission}, $A_{\rm 2ph}$ is the rate of the two photon decay and $C_{21}$ is the rate of the collisional de-excitation from the first excited state to the ground state, respectively. These rates can be written as
\begin{align}
  A_{\rm Ly\alpha} &= \frac{n_{2\rm p}}{n_2}A_{\rm 2p1s}
    =\frac{1}{1+r_{\rm 2s2p}}A_{\rm 2p1s} \label{eq:A_LyA},\\
  A_{\rm 2ph} &= \frac{n_{2\rm s}}{n_2}A_{\rm 2s1s}
    =\frac{r_{\rm 2s2p}}{1+r_{\rm 2s2p}}A_{\rm 2s1s}. \label{eq:A_2ph}
\end{align}
where {$A_{\rm 2p1s}=6.25\times 10^8\:{\rm s^{-1}}, A_{\rm 2s1s} = 8.25
\:{\rm s^{-1}}$} and $r_{\rm 2s2p}$ are the Einstein coefficients for the transition ${\rm 2p}\to{\rm 1s}$, ${\rm 2s}\to {\rm 1s}$ and the ratio between the number densities in {the 2s and the 2p states, respectively}. 
{Due to the very short timescale to achieve the equilibrium among the number of particles in the 1s, 2s and 2p states, the transition rates from/into the 2s state can be considered to balance and therefore the ratio $r_{\rm 2s2p}$ can be calculated as}
\begin{equation}
    r_{\rm 2s2p}=\frac{C_{\rm 2p2s}}{C_{\rm 2s2p}+A_{\rm 2s1s}}.
\end{equation}
where $C_{\rm 2s2p}$ and $C_{\rm 2p2s}$ are the rates of the transition 2s$\to$ 2p and 2p$\to$2s by collisions, respectively. 
{They are written as 
 \begin{equation}
    C_{\rm 2s2p} = 5.31\times10^{-4} \qty(\frac{n_{\rm e}}{1\mathrm{~cm}^{-3}}) \mathrm{~s}^{-1},
  \end{equation}
  \begin{equation}
    C_{\rm 2p2s} = C_{\rm 2s2p} \frac{g_{\rm 2s}}{g_{\rm 2p}}
    = 1.77\times10^{-4} \qty(\frac{n_{\rm e}}{1\mathrm{~cm}^{-3}}) \mathrm{~s}^{-1}.
  \end{equation}
  where $g_{\rm 2s}=2$ and $g_{\rm 2p}=6$ are the statistical weight of the 2s state and the 2p state and $n_e$ is the electron number density, respectively (see \citealt{seaton1955cross}).} 

The rate of the collisional de-excitation \citep{omukai2001primordial,inayoshi2016hyper}, $C_{21}$, is given by 
  \begin{equation}
    C_{21}=\gamma_{21}({\rm e}) n_{\rm e} + \gamma_{21}({\rm H}) n_{\rm H} \label{eq:C_21},
  \end{equation}
where $n_{\rm H}$ is the hydrogen number density, $\gamma_{21}({\rm e})$ and $\gamma_{21}({\rm H})$ are the de-excitation coefficients for the collisions with electrons and hydrogen atoms, respectively, which are given by substituting $(u,l)=(2,1)$ into the formulae \citep{1995eabs.book.....S,drawin1969influence} as below:
  \begin{equation}
    \gamma_{u l}( {\rm e})=10^{-8}\left(\frac{l^{2}}{u^{2}-l^{2}}\right)^{3 / 2} \frac{l^{3}}{u^{2}} \alpha_{l u} \frac{\sqrt{\beta(\beta+1)}}{\beta+\chi_{l u}}, \quad \beta=\frac{h\left(v_{l}-v_{u}\right)}{k T},
  \end{equation}
  \begin{align}
    \gamma_{ul}({\rm H})
     =7.86 \times 10^{-15}\left(\frac{l}{u}\right)^{2}\left(\frac{1}{l^{2}}-\frac{1}{u^{2}}\right)^{-2} f_{l u} T^{1 / 2}\notag \\
    \times \frac{1+1.27 \times 10^{-5}\left(1 / l^{2}-1 / u^{2}\right)^{-1} T}{1+4.76 \times 10^{-17}\left(1 / l^{2}-1 / u^{2}\right)^{-2} T^{2}}.
    \label{eq:gamma}
  \end{align}
where the coefficients are $\alpha_{12}=24$ {and} $\chi_{12}=0.28$.

\subsection{Numerical setup}

We solve discretized Eq.\ref{eq:BasicEq} based on the finite volume method by using the alternating-direction implicit (ADI) method.
 We set the gas temperature, 
the ionization degree, 
the mean molecular weight
to be $T =10^4 {\rm K}$, $x_e = n_e/n_{\rm HI}=10^{-4}$ and $\mu=1$, respectively
{(see also \citealt{inayoshi2016hyper,sakurai2016hyper}).} 
The profiles of the gas density and velocity are calculated by solving the fluid equation for the Bondi accretion without radiation effects, using the 11 parameter sets of $(n_\infty,M_{\rm BH})$ ({all parameter sets are shown in Table \ref{table:1} of the model "Bondi"})
 The Bondi accretion rate, $\dot M_{\rm B}$, is given by
    \begin{equation}
      \dot{M}_{\rm B}
      \approx4\times 10^3
    \left(\frac{{L}_{\rm Edd}}{c^2} \right)
      \left(\frac{M_{\rm BH}}{10^4M_\odot}
      \right)
      \left(\frac{n_\infty}{10^5\:{\rm cm^3}}\right)
      \left(\frac{T}{10^4{\rm K}}\right)^{-3/2}.\label{eq:MdotB}
    \end{equation}
where 
$L_{\rm Edd}$ is the Eddington luminosity for the Thomson scattering.

The total optical depth at the Ly$\alpha$ line center, measured over the radial line of the computational domain, $\tau_0$, can be approximately written as 
    \begin{align}
      \tau_0 \approx 4.3 \times 10^{11} \left(\frac{n_\infty}{10^5\:{\rm cm^3}}\right) \left(\frac{M_{\rm BH}}{10^4M_\odot}\right)\left(\frac{T}{10^4{\rm K}}\right)^{-3/2}
        \qty(\frac{r_{\min}}{10^{-3}r_{\rm B}})^{-1/2} ,\label{eq:tau0_Bondi}
    \end{align}
{assuming that the number density distirbution $n(r)\propto r^{-3/2}$ and the outer boundary of the radial integral of $\sigma_0 n(r)$ is set to infinity,}
where $r_{\rm B}$ is the Bondi radius which {depends} on $M_{\rm BH}$,
$T$, and $r_{\rm min}$ which 
is the innermost radius of the computational domain defined below. 

{The radii of the inner and outer boundaries of the computational domain are 
{$r_{\rm min}=10^{-3}r_{\rm B}$ and {$r_{\rm max}=10^{-1}r_{\rm B}$}, respectively. 
The innermost radius is the same as in \cite{inayoshi2016hyper,sakurai2016hyper}.
The number of radial grid cells is set to be $N=800$}}. 
{The radial grid is set so that the sequence of the radial length of cells becomes the geometrical sequence with the initial value 
$\Delta r_0 = 10^{-7}r_{\rm min}$ and the common ratio $\epsilon$.
 Here, 
$\Delta r_0$ corresponds to the size 
of the finest (innermost) grid cell and the common ratio $\epsilon\approx 1.02131$ satisfies the equation below: 
\begin{align}
r_{\rm max}- r_{\rm min} = \Delta r_0 \frac{\epsilon^N-1}{\epsilon-1}.
\end{align}
Such a grid design is convenient since we can control the size 
of the finest grid cell without changing $r_{\rm min},r_{\rm max}$ and $N$. {Note that it is irrelevant that the simulation domain does not include the Bondi radius because both the gravitational and radiation forces at such a distant region are small and their impacts  on the flow} are negligible (see also Fig.\ref{rfprofile}).}

The grid for the normalized frequency spans over the range of $[x_{\rm min},x_{\rm max}]$,
where we set 
$x_{\min}=-2(a\tau_0)^{1/3}$, and $x_{\max}=5(a\tau_0)^{1/3}$.
The grid interval $\Delta x_0$ is set to unity within the range $[-10,10]$, and $\Delta x_0 = (x_{\rm max}-x_{\rm min})/100$ {otherwise}, considering the significant and steep change in the line profile near the line center.

The situation where 
Ly$\alpha$ photons are entering the computational domain, passing
through the inner boundary at the rate of $L_{{\rm Ly}\alpha}$, is reproduced by setting the emissivity at the innermost cell as follows:
    \begin{equation}
      \eta_{x_0}(r_0) = \frac{L_{\rm Ly\alpha}}{4\pi} \frac{1}{\Delta V_0} \frac{H(x_0)}{\sqrt{\pi}}.\label{eq:eta}
    \end{equation}
where $\Delta V_0$ is the volume of the innermost cell. 
{The inner boundary condition of the spatial grid is set as the mirror boundary.}

Assuming the radiation is isotropic and can leak freely from outer boundary, the outer boundary condition can be described as 
    \begin{equation}
      F_0(r_{\max},x_0) = \frac{1}{2}cE_0(r_{\max},x_0).
    \end{equation}
{The radiation energy density at the frequency grid boundaries $x_0=x_{\min}$ and $x_{\max}$ is set to be zero at all radii.}

We investigate the two different {cases of accretion flow} named as the "slim" and "Eddington" {cases}.
In the "slim" {case}, the Ly$\alpha$ luminosity is given as 
    \begin{equation}
      L_{\rm Ly\alpha}= 0.2 \qty[1+\ln \qty(\dfrac{{\dot M_{\rm B}c^2}}{20 L_{\rm Edd}})] L_{\rm Edd},
    \label{eq.slim}
    \end{equation}
by assuming $10\%$ 
of the disk luminosity is converted into Ly$\alpha$ luminosity \citep{sakurai2016hyper} {and using the disk luminosity of the slim disk model from \cite{Watarai2000}}.
{It should be noted that the accretion rate of our models is typically $10^3$ times larger than the Eddington limit as seen in Eq.\ref{eq:MdotB}.}
In the "Eddington" case, we employ $L_{\rm Ly\alpha}= 0.1L_{\rm Edd}$ \citep{inayoshi2016hyper}. {We will primarily focus on discussing the result of the "slim" {case} while the "Eddington" {case} will also be used to compare with the previous studies.}

\begin{table}
    \centering
    \begin{tabular}{|c|c|c||c|c|c|}
    \hline 
    model & $M_{BH}[M_\odot]$ & $n_{\infty}[{\rm cm^{-3}}]$ & $\tau_0$ &$M_F$ \\
    \hline 
 Bondi     &  $10^4$        & $10^4$    &  $4.2\times 10^{10}$ &  110   \\
 Bondi     &  $10^4$        & $10^5$    &  $4.2\times 10^{11}$ &  110   \\
 Bondi     &  $10^4$        & $10^6$    &  $4.2\times 10^{12}$ &   120 \\
 Bondi     &  $10^{4.5}$        & $10^6$    &  $1.3\times 10^{13}$ &  130   \\
 Bondi     &  $10^5$        & $10^4$    &  $4.2\times 10^{11}$ &   100  \\
 Bondi     &  $10^5$        & $10^5$    &  $4.2\times 10^{12}$ &   120  \\

 Bondi     &  $10^5$        & $10^6$    &  $4.2\times 10^{13}$ &  140  \\
  Bondi     &  $10^{5.5}$        & $10^{5}$  &  $1.3\times 10^{13}$ & 130    \\
 Bondi     &  $10^6$        & $10^4$    &  $4.2\times 10^{12}$ &   140  \\
 Bondi     &  $10^6$        & $10^5$    &  $4.2\times 10^{13}$ &  140   \\
 Bondi     &  $10^6$        & $10^6$    &  $4.2\times 10^{14}$ &  150  \\
 hom. sphere     &  N/A   & $10^9$ & $10^{10}$& 180\\
 hom. sphere     &  N/A   & $10^9$ & $10^{11}$& 180\\
 hom. sphere     &  N/A   & $10^9$ & $10^{12}$& 180\\
 hom. sphere     &  N/A   & $10^9$ & $10^{13}$& 180\\
 \hline
    \end{tabular}
    \caption{Model parameters and results. 
    {The model parameters $\dot M_{\rm BH}$ and $n_\infty$ are the BH mass and the density at infinity, respectively, $\tau_0$ is the total optical depth measured over the computational domain along the radial direction and $M_F$ is the resulting force multiplier. The model name "Bondi" indicates the model where the Bondi-like fluid field is used and "hom. sphere" means the model where the homogeneous and static medium is used and thus the system in this model predominantly depends on the optical depth $\tau_0$ and weakly depends on density $n_\infty$ (see Fig.\ref{fig4} for more detail)}}
    \label{table:1}
\end{table}

%% file: result_E.tex
\subsection{Radial profile of Ly$\alpha$ radiation force}

\begin{figure}
      \centering
      \includegraphics[width=\columnwidth]{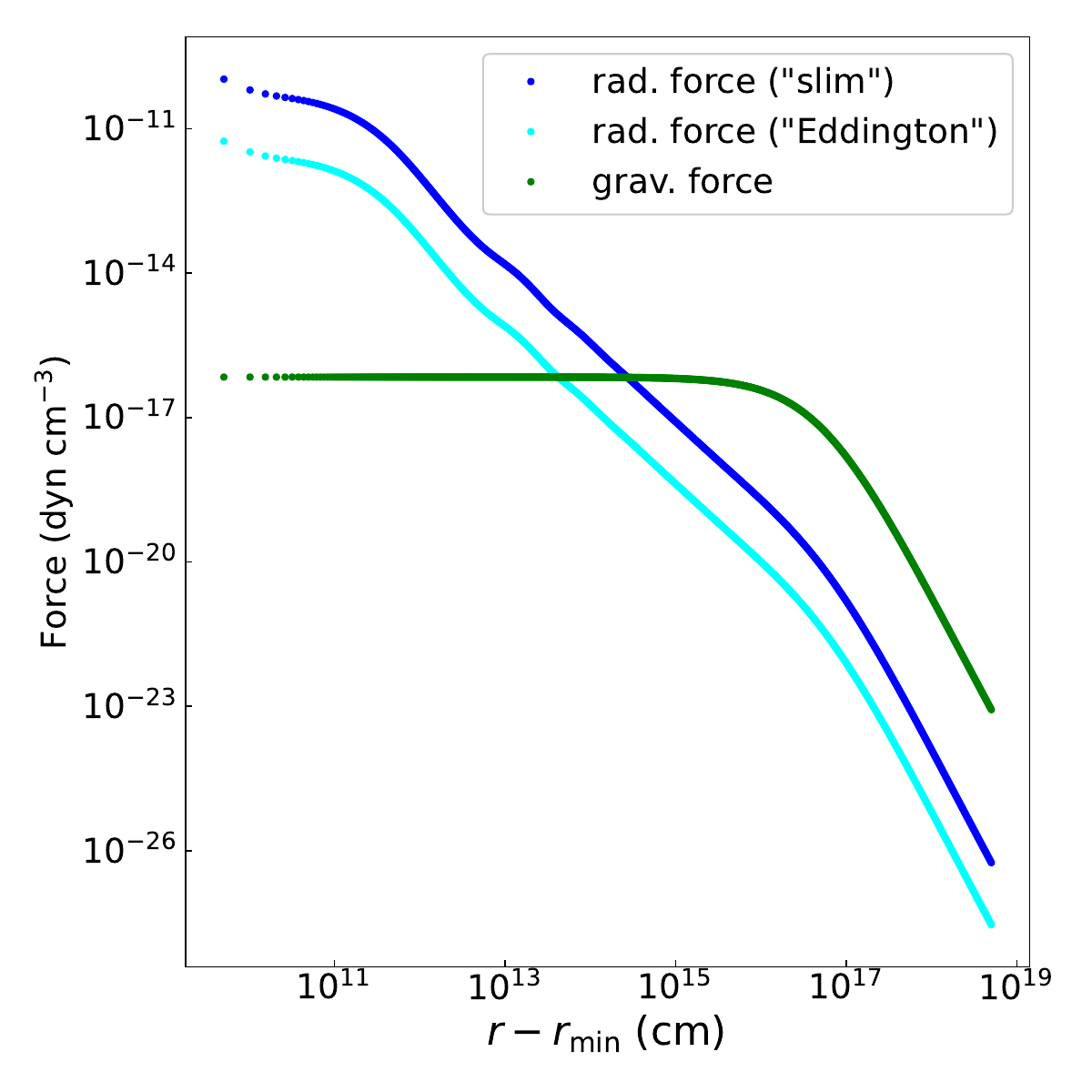}
      \caption{Radial profiles of the Ly$\alpha$ radiation force and gravity (green) for $M_{\rm BH} = 10^{5.5} M_{\odot},n_{\infty}=10^{5} {\rm ~cm^{-3}}.$
      The horizontal axis shows the radial coordinate measured from $r=r_{\rm min}$, while the vertical axis shows the force exerted on the gas per unit volume in cgs unit. 
      {The blue (cyan) points show the radiation force in the "slim" ("Eddington", respectively) model.}
      \alttext{Alt text: Comparison between the gravity and radiation forces in the radial profile.}
      }
      \label{rfprofile}
    \end{figure}

In Fig.\ref{rfprofile} we present radial profiles of the Ly$\alpha$ radiation force per unit volume,
    \begin{align}
      f_{\rm rad} = -\int \frac{1}{3}\pdv{E_{x_0}}{r} dx_0,
      \label{eq:f_rad}
    \end{align} 
along with gravity for the case with $M_{\rm BH} = 10^{5.5} M_{\odot}$ and $n_{\infty}=10^5 {\rm ~cm^{-3}}$ both for the "slim" and "Eddington" {cases}.

It is found from the figure that the Ly$\alpha$ radiation force is higher in the "slim" {case} than in the "Eddington" {case} since the Ly$\alpha$ photons are emitted more efficiently in the "slim" case. 

In both cases, within the region $r-r_{\min}\lesssim 10^{12} {\rm~cm}${, where $r_{\rm min}\simeq 6.2\times 10^{16} \:{\rm cm}$},
the radiation force is significantly enhanced due to numerous photon scatterings, exceeding gravity by 
approximately $10^6$ times. 
However, as the radius increases,
the radiation force decreases monotonically and drastically in both {cases}, {falling below gravity for $r-r_{\min}\gtrsim 5\times 10^{16} {\rm~cm}$.
This decline is attributed to the two-photon decay process.
Ly$\alpha$ photons are destroyed in the vicinity of the inner boundary since the photon destruction length, which is the distance a photon travels before it is destroyed
via the two photon decay, is very small $\sim 10^{12}{\rm \: cm}$ (see also Eq. \ref{eq:Dr_pd} in Appendix \ref{Appendix:comparison})}.
Thus, the number of photons decreases dramatically with increasing $r-r_{\rm min}$, leading to a reduction in the radiation force.

The observed trend, where the radiation force surpasses gravity in the innermost region and decreases with increasing radius, is consistent across all parameter sets. For computational domains with smaller optical depths, the radius (normalized by the Bondi radius) of the region where the radiation force exceeds gravity is larger.
In the case with the smallest optical depth in this study $(M_{\rm BH},n_{\infty})=(10^4{M_\odot},10^4{\rm cm^{-3}})$, the radiation force in the "slim" {case} overwhelms gravity throughout the entire computational domain.

\subsection{Force multiplier}\label{sec3.2}

\begin{figure}[!tp] 
      \centering
      \includegraphics[width=\columnwidth]{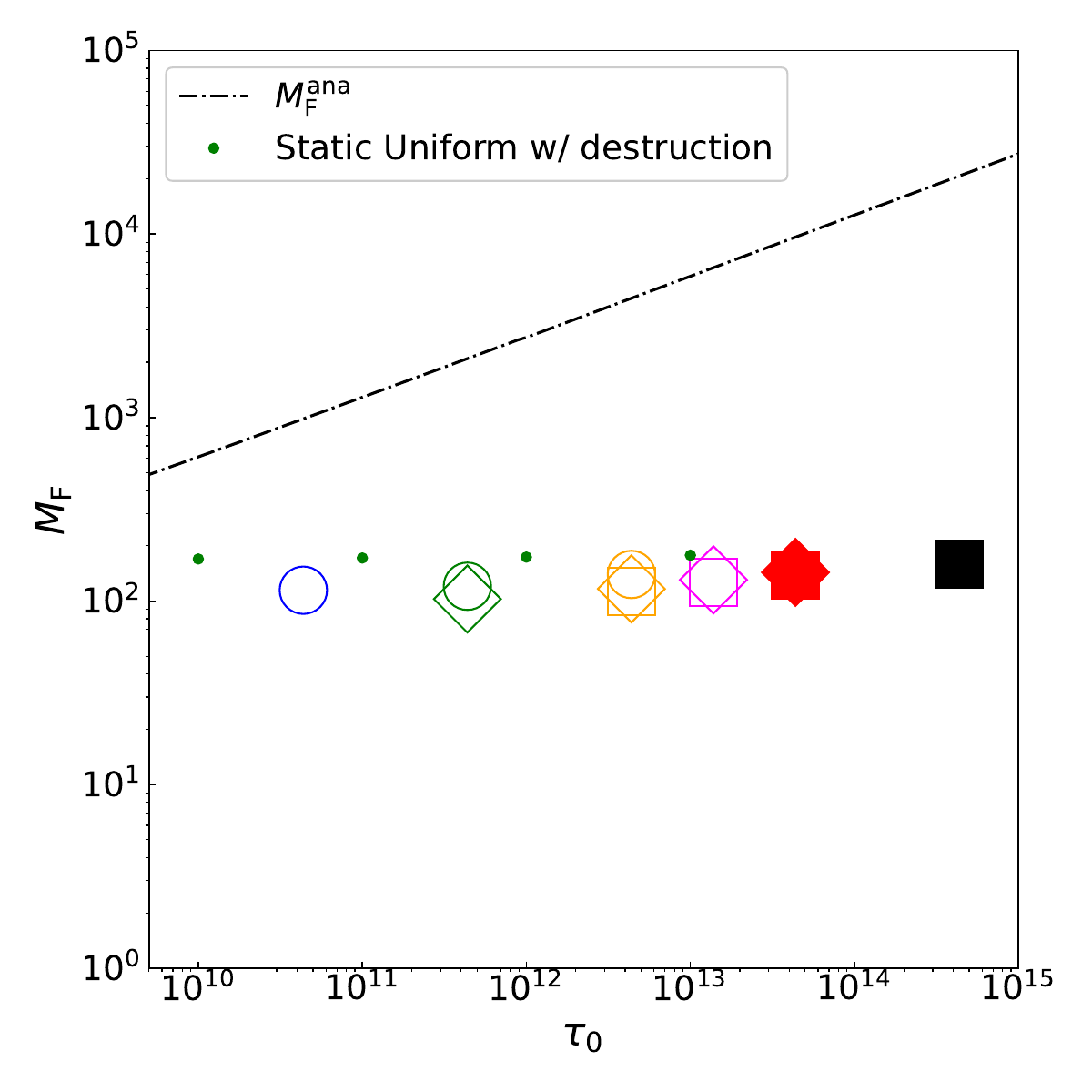}
      \caption{
      The relationship between the optical depth of the entire system at the Ly$\alpha$ line center and the force multiplier of the Ly$\alpha$ radiation. 
      {The same symbols represent models with the same number density $n_\infty$ (the model with $n_\infty = 10^4,10^5,10^6 \:{\rm cm^{-3}}$ corresponding to the circle, diamond, square symbols,respectively) and the same colors of the symbols indicate the models with the same optical depth through the system, $\tau_0$}
      Whether or not the symbol is filled with color indicates whether the relation {$F_{\rm rad}<F_{\rm grav}$} (refer also to the description around Eq.\ref{eq:F_grav}) is satisfied or not. The black dashed line represents the analytical solution for the case of static homogeneous hydrogen gas, $M^{\rm ana}_F$, given by Eq.\ref{eq:MFassymp}. The green points show the results of the computation of static homogeneous gas with destruction processes.
      \alttext{Alt text: The resulting force multiplier from each model. The force multiplier is nearly constant at around 100 in all our models, with optical depths $\tau_0$ ranging from $10^{10}$ to $10^{15}$.}
      }
      \label{tauMprofile}
    \end{figure}

Fig.\ref{tauMprofile} depicts the relationship between the total optical depth at the Ly$\alpha$ line center across the entire computational domain and the force multiplier {$M_{\rm F}$\footnote{{It is worth noting that the force multiplier is sometimes defined as the ratio between the actual radiation force and the radiation force considering only Thomson opacity \citep{CAK1975}. However, the definition used here is different from it.}}, which is defined as: 
  \begin{equation}
    M_{\rm F} = \frac{c}{L_{\rm Ly\alpha}}\int f_{\rm rad}dV.
    \label{eq:M_F}
  \end{equation}
  }
We can observe that the force multiplier remains almost independent of the optical depth, maintaining a constant value of $M_{\rm F} \simeq 130$. 
{
The reason for this 
is primarily attributed to the effect of the two-photon decay, rather than the Doppler shift caused by the velocity and its gradient.}

To illustrate the reason, we examine the following two cases. The first is the case of a static, homogeneous hydrogen medium where \ly photons are assumed not to be destroyed. 
In this case, $M_{\rm F}$ can be analytically described\footnote{We use the definition ${\rm arctanh}(x)=\frac{1}{2}\ln{\left(\frac{1+x}{1-x}\right)}.$} as \citep{tomaselli2021lyman}
    \begin{align}
      M_{\rm F}^{\rm ana}(\tau_0) = \frac{4}{\pi}\sqrt{\frac{2}{3}}\int_{-\infty}^{\infty}{\rm arctanh}(e^{-\pi|\sigma(x)|/\tau_0}){dx},\label{eq:MFana}
    \end{align}
where $\sigma(x)=\int_0^x \sqrt{2/3}/H(y,a)dy$
In the limit of a large optical depth, this equation can be approximated as:
    \begin{equation}
      M_{\rm F}^{\rm ana}(\tau_0) \approx 3.51(a\tau_0)^{1/3} = 610\qty(\frac{\tau_0}{10^{10}})^{1/3}\left(\frac{T}{10^4{\rm K}}\right)^{-1/6},\label{eq:MFassymp}
    \end{equation}
which is also represented in Fig.\ref{tauMprofile} by the black dot-dashed line.
{The second is the case where destruction of \ly photons are taken into consideration in a static homogeneous hydrogen gas. The results of numerical calculation are shown by the green points. It is observed that the resulting $M_{\rm F}$ is considerably smaller than that of the first case, closely agreeing with the outcomes of our present study}.
This indicates that the radiation force diminishes and $M_{\rm F}$ becomes constant with $\tau_0$, not due to the velocity gradient, but owing to photon destruction.
{This fact can be understood as follows: 
\ly photons undergo numerous scattering and vanish when traveling the distance equivalent to the mean free path for the photon destruction process.
Consequently, the \ly radiation force effectively work only within the extent that \ly photons can penetrate, and it becomes approximately constant regardless of $\tau_0$, even when the optical depth of the system is considerably large.
Since the region penetrated by \ly photons is extremely narrow, the velocity change within the region is small. Therefore, the effect of the velocity field on $M_{\rm F}$ is negligibly small
(see Appendix \ref{Appendix:comparison} for more details). }
Note that the constancy of the force multiplier 
holds true only when the gas number density $n_\infty$ is between $10^8{\: \rm cm^{-3}}$ and $10^{12}{\:\rm cm^{-3}}$ as mentioned in Appendix \ref{Appendix:comparison}. It is also noteworthy that 
the properties of $M_{\rm F}$ discussed in this subsection are independent of $L_{\rm Ly\alpha}$. This is because the radiation force, $f_{\rm rad}$, is proportional to $L_{\rm Ly\alpha}$ via $E_{x_0}$
(see Eq. \ref{eq:f_rad}).

\subsection{Comparison between Ly$\alpha$ radiation force and gravity
}\label{sec:comparison}
\begin{figure}[!pt]
      \centering
      \includegraphics[width=\columnwidth]{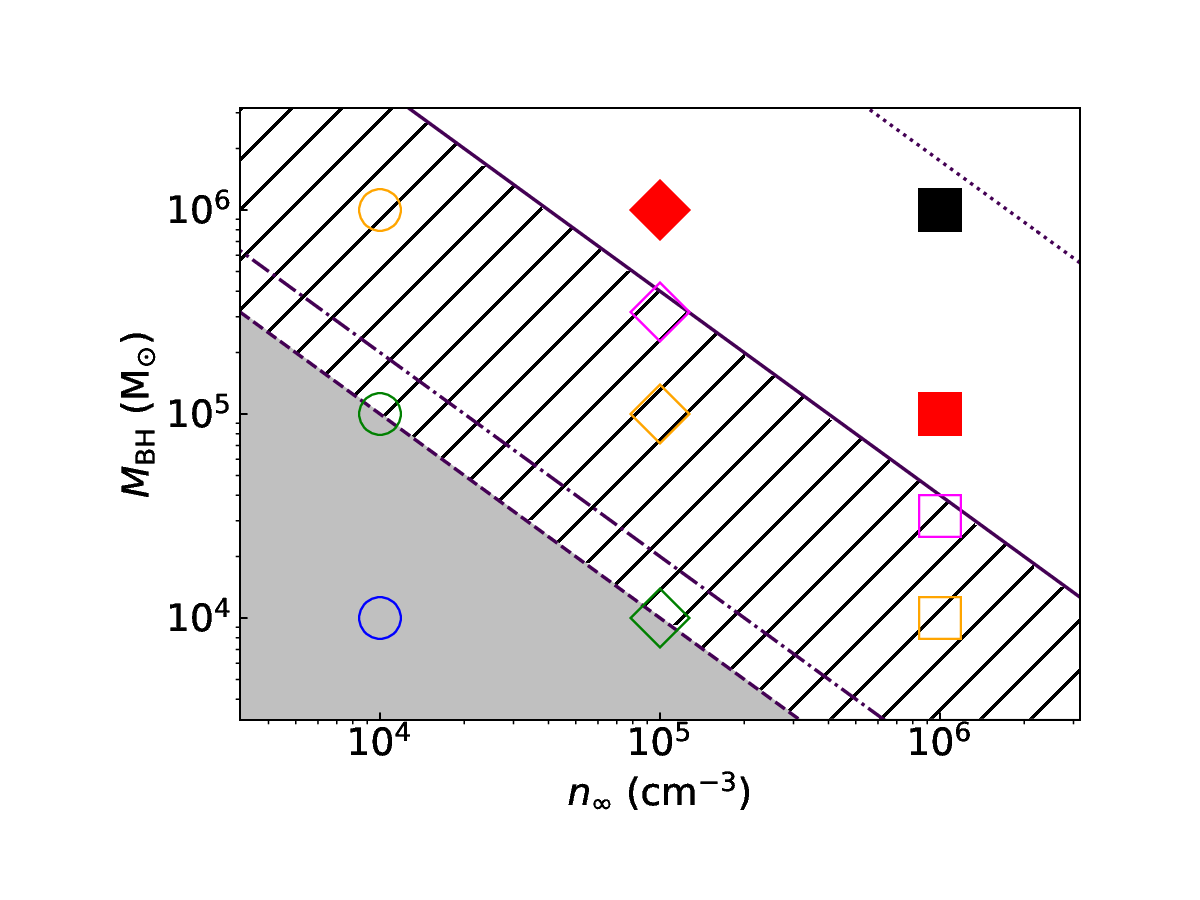}
      \caption{
      The feasibility of the supercritical accretion in all models within the "slim" category. {Each symbol shows the result of the computation for a specific model, with its color corresponding to the model with the different parameter set $(M_{\rm BH},n_\infty)$.} Filled symbols show the models where gravity prevails over radiation, {expressed as $F_{\rm grav}>F_{\rm rad}$,} while unfilled symbols show the opposite cases. The lines delineate the boundaries where the supercritical accretion can be achieved, with $(n_{\infty}/10^5{\rm cm^{-3}})(M_{{\rm BH}}/10^4 M_\odot)=$
      1(dashed), 2(dot-dashed), {40}(solid) {and 1500(dotted)}. {The dashed line} corresponds to the results in \protect\cite{inayoshi2016hyper,sakurai2016hyper}. {The dot-dashed and solid lines correspond to the results of the simple formula Eq.\ref{eq:F_ratioNum} with $r_{\rm min}=10^{-3}r_{\rm B}$, applied for the "Eddington" and "slim" models, respectively. The dotted line corresponds to that with $r_{\rm min}=r_{\rm B}$, applied for the "slim" model.
      \alttext{Alt text: The graph shows the area of models where super-Eddington accretion becomes feasible in the $n_\infty-M_{\rm BH}$ diagram. We calculated 11 models, and only in 3 models does gravity overwhelm the radiation force in the 'slim' model.}
      }}
      \label{tauMdiagram}
    \end{figure}

Fig.\ref{tauMdiagram} summarizes the possibility of the supercritical accretion on the $n_\infty$-$M_{\rm BH}$ plane for the "slim" models. {The filled symbols show the models where the super-Eddington accretion is possible}
as $F_{\rm grav}>F_{\rm rad}$.
Here, $F_{\rm grav}$ and $F_{{\rm rad}}$ are gravitational and radiation forces
integrated across the entire simulation domain, 
defined as follows:
\begin{align}
    F_{\rm grav} &= 4\pi\int_{r_{\rm min}}^{r_{\rm max}} {G\rho M_{\rm BH}} dr,\label{eq:F_grav}
    \\
    F_{\rm rad} 
    &=\int_{r_{\rm min}}^{r_{\rm max}} f_{\rm rad} 4\pi r^2 dr
    =\frac{M_F}{c} L_{{\rm Ly}\alpha}.
    \label{eq:F_rad}
\end{align}
{The radiation force $F_{\rm rad}$ is proportional to the force multiplier $M_F$ and the Ly$\alpha$ luminosity. The latter is given by the product of the Eddington luminosity and a factor that is either fixed to $0.1$ for the "Eddington" case or depends on $n_\infty M_{\rm BH}$ from Eqs.\ref{eq:MdotB} and \ref{eq.slim} for the "slim" case.}

This figure shows that the \ly radiation force elevates the lower limit of $n_\infty M_{\rm BH}$ for realizing super-Eddington accretion.
{The grey area represents the parameter range where the super-Eddington accretion is unattainable even in the absence of the Ly$\alpha$ radiation force \citep{inayoshi2016hyper,sakurai2016hyper}\footnote{{Note that \cite{Toyouchi_et_al2019} derives the relation where the required value for $n_\infty M_{\rm BH}$ is one order of magnitude larger than the previous studies by considering the pressure balance on the boundary of HII region, but without the Ly$\alpha$ radiation.}}. 
By incorporating the Ly$\alpha$ radiation force, 
the region of non-supercritical accretion expands into the shaded area.
Our calculations confirm that the seven models within the shadow region are incapable of supporting super-Eddington accretion.

The boundary of the shaded area and the unshaded region (super-Eddington are) can be derived} through analytical considerations.
{Assuming the density profile of the Bondi accretion at the inner limit, $n(r)=n_{\infty}\left(r/r_{\rm B}\right)^{-3/2}$, {the graviational force $F_{\rm grav}$ (Eq.\ref{eq:F_grav}) can be written as }
\begin{align}
    {
    F_{\rm grav}\simeq 4.2\frac{L_{\rm Edd}}{c}\left(\frac{n_{\infty}}{10^5{\rm cm^{-3}}}\right)\left(\frac{M_{{\rm BH}}}{10^4M_\odot}\right)\left(\frac{T}{10^4{\rm\: K}}\right)^{-1}\left(\frac{r_{\rm min}}{10^{-3}r_{\rm B}}\right) ^{-1/2}
    }
\end{align}
{and thus }the ratio of $F_{\rm rad}$ and $F_{\rm grav}$} can be expressed as
{
    \begin{align}
      \frac{F_{\rm rad}}{F_{\rm grav}} \sim 2 
      &\left(\frac{n_\infty}{10^5{\rm cm^{-3}}}\right)^{-1}
      \left(\frac{M_{\rm BH}}{10^4M_\odot}\right)^{-1}\nonumber\\ \times
      &\qty(\frac{M_{\rm F}}{100}) \qty(\frac{L_{\rm Ly\alpha}}{0.1L_{\rm Edd}})
             \qty(\frac{r_{\min}}{10^{-3}r_{\rm B}})^{1/2}\left(\frac{T}{10^4{\rm K}}\right).\label{eq:F_ratioNum}
    \end{align}
    }
The force multiplier, $M_{\rm F}$, is approximately $100$ in the parameter sets employed in our study (see also fig.\ref{tauMprofile}), 
allowing us to deduce $(n_{\infty}/10^5{\rm cm^{-3}})(M_{{\rm BH}}/10^4 M_\odot) ={40}$
from $F_{\rm rad}=F_{\rm grav}$ {for the "slim" case}.
Consequently, super-Eddington accretion occurs when the gas density and the BH mass satisfy the condition $(n_{\infty}/10^5{\rm cm^{-3}})(M_{{\rm BH}}/10^4 M_\odot)\gtrsim {40}$ for the "slim" case.
A similar argument yields the condition $(n_{\infty}/10^5{\rm cm^{-3}})(M_{{\rm BH}}/10^4 M_\odot) \gtrsim 2$ for the "Eddington" case.
Note that, in the "slim" case, the condition is more stringent because of the luminosity exceeding the Eddington limit, resulting in a stronger radiation force.

The reason why the relation depends only on $M_{BH}n_{\infty}$ even in the "slim" case is attributed to the fact that {the ratio $L_{{\rm Ly}\alpha}/L_{\rm Edd}$} depends only on {$\dot M_{\rm B}/(L_{\rm Edd}/c^2)$}, which is proportional to $n_{\infty}M_{\rm BH}$ {(see Eqs.\ref{eq:MdotB} and \ref{eq.slim})}. The boundary drawn under this assumption agrees with our results. 
It should be noted that the analytical description of $F_{\rm rad}/F_{\rm grav}$ includes the inner boundary radius, $r_{\rm min}$, which is fixed to $r_{\rm min}=10^{-3}r_{\rm B}$ in our model but nearly unknown in realistic cases. 
{If the inner boundary radius $r_{\rm min}$, corresponding to the HII region redius, increases, the gravitational force becomes weaker in the innermost region, facilitating the dominance of the radiation force over gravity.} {According to the simulation of the super-Eddington flow in \cite{inayoshi2016hyper}, the size of the HII region in the early phase is similar to the Bondi radius. At this stage, ${\rm Ly}\alpha$ radiation becomes more important. Under the condition of $r_{\rm min}\sim r_{\rm B}$, the required value of $n_{\infty,5}M_{{\rm BH},4}$ to realize the super-Eddington accretion becomes about thirty times larger than that with $r_{\rm min}=10^{-3} r_{\rm B}$ and reaches $\sim 1500$ for the "slim" case.}

%% file: summary_and_discussions.tex
To assess the impact of the Ly$\alpha$ radiation force on Bondi-like super-Eddington accretion flows, {as} previously discussed in \cite{inayoshi2016hyper,sakurai2016hyper}, we have calculated the {radiation field} of Ly$\alpha$ photons {using the diffusion approximation. Our calculations take into account the destruction processes, such} as two photon decay and collisional de-excitation. {We have explored the parameter range of $10^4\le M_{\rm BH}\:[M_\odot]\le 10^6$ and $10^4  \le n_\infty\:[{\rm cm^{-3}}]\le 10^6$.}
Our model supposes that the computational domain is filled with an isothermal Bondi flow of HI gas and that \ly photons are injected through the inner boundary, considering that
the gas located closer to the center than the inner boundary is ionized by the UV photons emitted from the black hole accretion disk.
Below we summarize our key findings:

\begin{itemize}
\item 
We have found that the Ly$\alpha$ radiation force becomes much stronger than the gravity in the immediate vicinity of the inner boundary.
However, as distance increases, the influence of the radiation force diminishes, and gravity becomes the predominant force at larger distances. 
This implies that the Ly$\alpha$ radiation force effectively hinders inflow motion solely within the narrow region of atomic hydrogen (HI) gas immediately outside the ionized HII region.

\item 
The pronounced strength of the \ly radiation force is attributed to resonance scatterings. Despite undergoing numerous scatterings, \ly photons are ultimately annihilated by destructive processes such as two-photon decay and collisional de-excitation. Given that many \ly photons penetrate slightly from the inner boundary and are subsequently destroyed, the \ly radiation force experiences enhancement only near the innermost edge of the HI region. The force multiplier remains constant around 130 and does not increase with the optical thickness of the system.

\item
The strong \ly radiation force necessitates a higher value of $n_\infty M_{\rm BH}$ to realize super-Eddington accretion. 
Specifically, when the luminosity of the black hole accretion disk follows the prediction of the slim accretion disk, 
the condition for realizing super-Eddington accretion to is given by $(n_{\infty}/10^5 {\rm cm^{-3}} ) (M_{\rm BH}/10^4M_\odot) \gtrsim {40}$, in contrast to $(n_{\infty}/10^5 {\rm cm^{-3}} ) (M_{\rm BH}/10^4M_\odot) \gtrsim 1$ when the \ly radiation force is not taken into account \citep{inayoshi2016hyper}.
If the luminosity of the disk is limited by the Eddington luminosity, the condition is relaxed $(n_{\infty}/10^5 {\rm cm^{-3}} ) (M_{\rm BH}/10^4M_\odot) \gtrsim 2$.
\end{itemize}
Even if we relax the diffusion approximation employed in the present study, the main results would remain largely unaltered. 
The diffusion approximation is valid only if the mean free path of the photon remains considerably smaller than the dimensions of the system.
The Doppler effect induces deviations in the frequency of \ly photons from the line center value, leading to an enhancement of the mean free path.
Consequently, the accuracy of calculations diminishes as the velocity gradient increases.
Indeed, the diffusion approximation reproduces the spectrum of the Hubble flow successfully in cases with small velocity gradient, while it fails in cases with large velocity gradient
(see Appendix \ref{appendix:test} for more detail).
In the case of the Bondi accretion flow, treated in our study, most \ly photons are destructed before the velocity gradient becomes influential
(Appendix \ref{Appendix:comparison}).
Although some photons
manage to survive destruction and their mean free path can be be comparable to or even greater than the size of the system, the radiation force exerted by such photons is not significantly large.
In instances involving low density or pronounced velocity gradients, a relaxation of the diffusion approximation becomes imperative for more accurate modeling, as discussed in Appendix B.

{
It is important to note, as mentioned in Sec. \ref{sec:comparison}, that the size of HII regions, corresponding to $r_{\rm min}$, is not well-defined in realistic cases. 
In the super-Eddington simulations by Inayoshi et al. (2016), the HII region gradually shrinks over time and eventually vanishes, meaning that the size of the HII region can be arbitrarily small.
In more realistic cases, an accretion disk would form inside the centrifugal radius. High-energy radiation from the disk effectively ionizes the surrounding gas, creating an HII region. In this case, the size of the HII region would be comparable to or larger than the centrifugal radius, which is similar to or greater than $\sim 10^{-1}M_{{\rm BH},4}$ times the Bondi radius (see Eq.19 in \citealt{sugimura2018stunted}). 
Recall that the HII region we considered is much smaller than those estimates. 
As the size of the HII region, $r_{\rm min}$, decreases, the ratio $F_{\rm rad}/F_{\rm grav}$ also decreases, as shown in Eq. \ref{eq:F_ratioNum}, making super-Eddington accretion more likely. In this sense, our estimate of the effect of the Ly$\alpha$ radiation force is conservative. 
Further investigation into the size of the HII region and the impact of Ly$\alpha$ radiation is awaited, particularly through radiation hydrodynamic simulations that incorporate Ly$\alpha$ radiative transfer.
}

The destruction of \ly photons, as demonstrated in the present study, plays an important role in the early universe, but should be more effective in environments with higher metallicity.
Assuming that the gas flow setup in this study contains typical astronomical silicate dust, the ratio of the destruction rate by the two-photon decay and the collisional de-excitation to that by dust absorption can be estimated as {$\sim 0.5\:(T/10^4 {\rm K})^{-1/2}(Z/Z_\odot)^{-1}$}, where $Z/Z_\odot$ denotes the metallicity relative to solar abundance.
This implies that \ly photons are  more effectively eliminated by dust absorption in flows with higher metallicity.
It should be noted, however, that in environments with higher metallicity, while the radiation force due to \ly photons is attenuated, the radiation force due to dust absorption becomes more enhanced.

While the \ly radiation force is evaluated here for fixed distributions of the density, temperature, and velocity, exploring its time evolution is a crucial aspect for future studies.
The \ly radiation force can dynamically alter the flow structure, thereby influencing its impact on the gas. 
To unravel whether super-Eddington accretion occurs, it is essential to study the time evolution of radiation fields coupled with fluid dynamics. 
This necessitates the application of radiation hydrodynamics simulations.

{
Although the present study assumes spherical symmetry, it is important to consider potential multi-dimensional effects. In cases where the accretion flow exhibits anisotropy, \ly photons may escape preferentially along directions characterized by lower HI column density, thus reducing the impact of the \ly radiation force that works to prevent accretion. 

It has been reported that multidimensional effects can suppress the decrease in the mass accretion rate \citep{sugimura2018stunted,Takeo2018}. 
The anisotropy in these cases arises from the gas possessing angular momentum and the anisotropic radiation emitted by the super-Eddington accretion flow.
Moreover, {\cite{Takeuchi_et_al.2013}} have reported that the radiation Rayleigh-Taylor instability induces anisotropic structures resembling gas clumps {\cite[see also][]{Takeuchi_et_al.2014,Jiang2013}}. Undertaking multi-dimensional radiation hydrodynamics simulations emerges as a pivotal consideration for future research.
}  

%% file: appendix_velocitytest.tex
\section{Validity of diffusion approximation}\label{appendix:test}

\begin{figure}
       \centering
       \includegraphics[width=0.8\columnwidth]{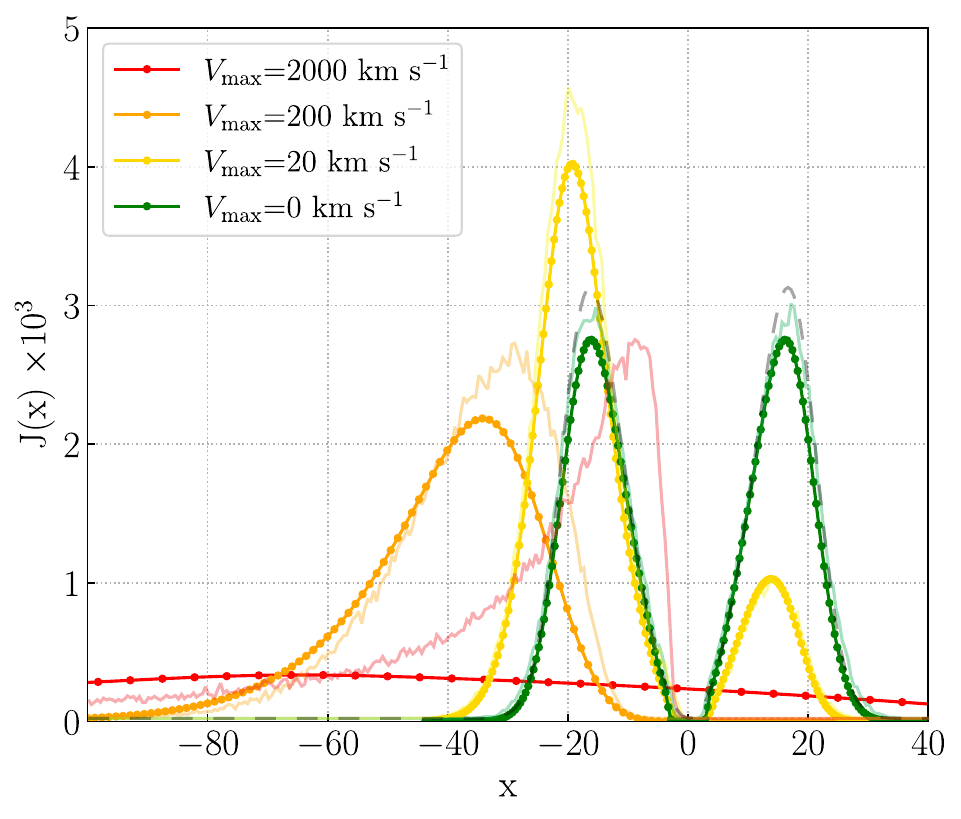}
       \caption{
        The emergent spectrum from homogeneous isothermal HI gas with the temperature $T=10^4 {\rm \: K}$, the Hubble flow velocity $v(r)$ in Eq. \ref{eq:v_hubble} and the column density $N_{\rm HI}=2\times 10^{20} {\:\rm cm^{-2}}$. Curves with different colors show results of the models with different maximum velocity $V_{\rm max}$. {The thin lines show the results of \cite{laursen2009lyalpha}, with their colors corresponding to the models with the same $V_{\rm max}$ as our results. The thin dashed black line is drawn by the analytical formula derived in \cite{dijkstra2006lyalpha}.
        \alttext{Alt text: Graphs comparing the emergent spectrum with the previous study across various velocity gradients. The horizontal axis shows the frequency normalized by the thermal width, ranging from -100 to 40. When the velocity gradient is large, the emergent profile shifts to the red and becomes asymmetric with respect to the line center.}}
        }
      \label{fig:hubbleflow}

    \end{figure}

 Our calculation includes the effect of velocity up to $O(v/c)$, which means the Doppler effect by the velocity and its gradient and the energy loss/gain by the work to/from the fluid are included. To check if these effects are solved correctly in our code, we compare our calculation with the Monte-Carlo calculation in the Hubble flow by \cite{laursen2009lyalpha}. The velocity field in the Hubble flow is given by
     \begin{equation}
      v(r)=V_{\max}\frac{r}{R_{\max}}
      \label{eq:v_hubble}
    \end{equation}
where the parameters $R_{\rm max}$ and $V_{\rm max}$ are the radius of the outer boundary and the maximum velocity, respectively. The gas is distributed homogeneously with a column density of $2\times10^{20} {\rm \: cm^{-2}}$ and is isothermal with the temperature of $10^4\:{\rm K}$. 
Fig.\ref{fig:hubbleflow} shows the frequency dependence of resulting radiation energy at $r=R_{\rm max}$, with the maximum velocity $V_{\rm max}$ varying from 0 ${\rm km\:s^{-1}}$ 
to ${\rm 2000\:km\:s^{-1}}$. 
Our calculations nicely reproduce the results of the previous study in the case of $V_{\rm max} = 0, 20, 200\:{\rm km\:s^{-1}}$. 
However, our results deviate from the correct answer
in the case of very large velocity,
$V_{\rm max} = 2000\:{\rm km\:s^{-1}}$.
This is because that
the diffusion approximation is invalid in a optically thin medium.
 Due to the Doppler shifting by the large velocity gradient, most photons immediately go out of the Ly$\alpha$ line center  and the mean free path of photons is dramatically increased.
Then, the diffusion approximation breaks down.
Our calculation cannot  correctly treat such a large velocity situation. 

The limit of the approximation can be estimated by comparing the energy shift by the resonance scattering with that by the velocity gradient. The shift by the velocity gradient is $\sim V_{\max}/v_{\rm th}$ 
while that by resonance scattering is $\sim (a\tau_0)^{1/3}\simeq 13 (N_{\rm  HI}/10^{20}{\rm cm^{-2}})^{1/3}$ (see Eq.\ref{eq:M_F_limit}) when photons traveling though the whole system, {where $v_{\rm th}\approx 13 \times (T/10^4{\rm K})^{1/2} {\rm km/s}$ and $a=4.7\times 10^{-4}\: T_4^{-1/2}$, respectively.}
Therefore, when the velocity is much greater than 
$\sim 200\:{\rm km\:s^{-1}}$, 
system becomes effectively very optically thin and the diffusion approximation is no longer valid. This argument is consistent with the results of our test calculations.
{
It should be noted that the diffusion approximation remains valid in the present study, even when the velocity reaches $\sim 1000 {\:\rm km/s}$ in the innermost region of the Bondi flow, since the column density of the system is much greater than in this test calculation. Consequently, the acceptable maximum velocity to maintain the validity of the diffusion approximation is also $\gtrsim 1000 {\:\rm km/s}$.
}

%% file: appendix_compare_homogeneous.tex
\section{Detailed comparison with static homogeneous gas with destrcution process}\label{Appendix:comparison}

\begin{figure}[!tp]
      \centering
      \includegraphics[width=\columnwidth]{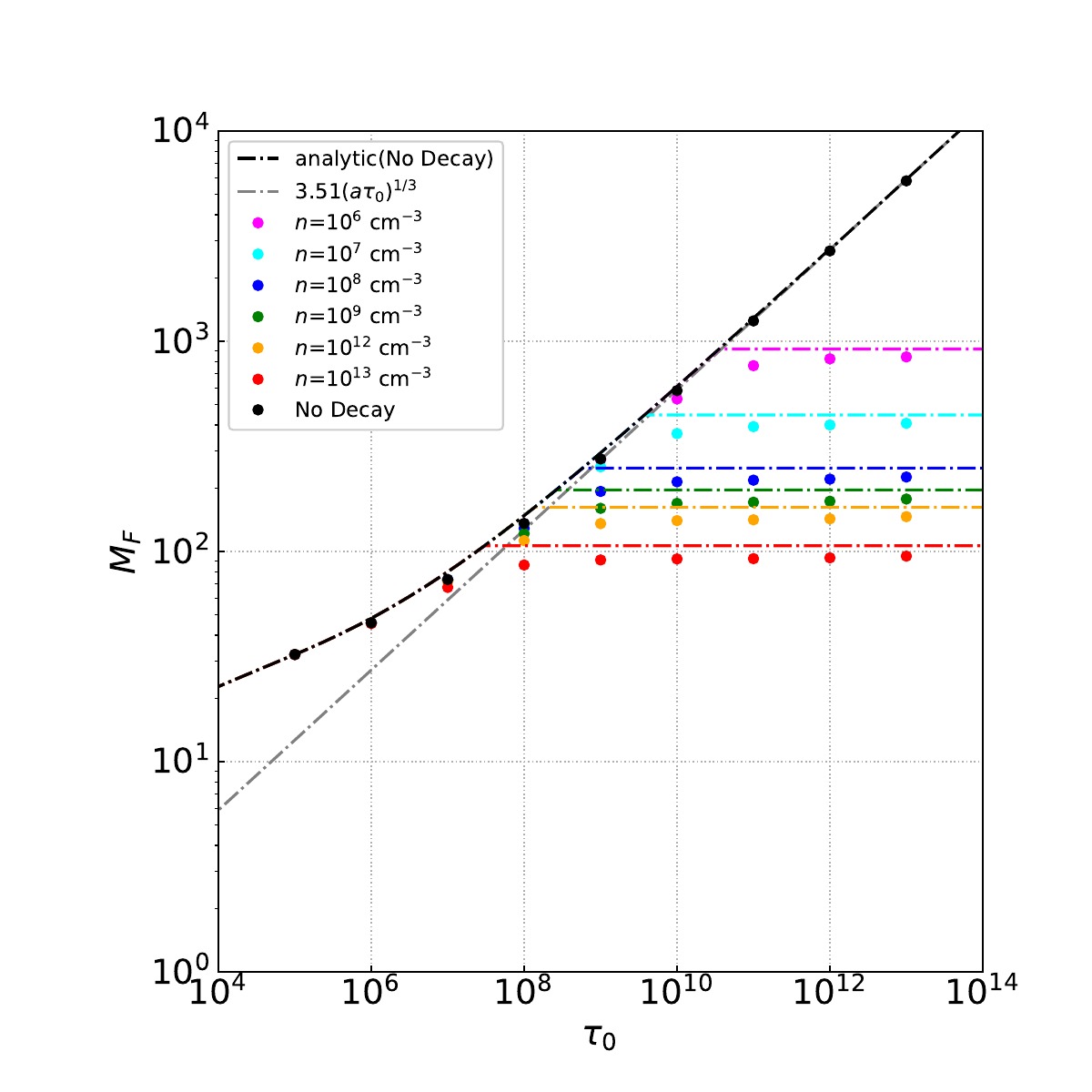}
    \caption{
    The force multiplier in a static homogeneous gas with the destruction process. Points in various colors, excluding the black ones, represent model results with the destruction process at different number densities, denoted as $n$.
    The black points depict the outcomes of calculations for a static homogeneous gas without the destruction process.
    The black {dot-dashed} line represents the analytical solution of the homogeneous model without the destruction process, given in Eq.(\ref{eq:MFana}) , while the grey {dot-dashed} line presents its approximated formula, proportional to $\tau_0^{1/3}$. The dot-dashed horizontal lines in various colors indicate the saturated values of the force multiplier at different number densities, as given by Eq.\ref{eq:MFana_annihilation}.
    \alttext{Alt text: Graph showing the relationship between the force multiplier and optical depth in static homogeneous gas with various densities. The number density ranges from $10^6$ to $10^{13}$ per cubic centimeter. As the number density increases, the terminal force multiplier also increases.}
    }
      \label{fig4}
    \end{figure}

We discuss the reason why the force multiplier is almost constant 
regardless of optical depth
when the destruction process is effective. 
Fig.\ref{fig4} shows the force multiplier in the static spherical homogeneous hydrogen gas with $T=10^4\:{\rm K}$, $x_e = 10^{-4}$ and with the destruction process and with the different $n_\infty$ as a function of the optical depth $\tau_0$. We can see clearly, in this figure, the saturation of the force multiplier, and also that the terminal values depend on the number density, $n_\infty$. 

Here, we confirm this trend through a simple analysis. 
The destruction of Ly$\alpha$ photons  occurs
when the number of scatterings approximately reaches $\sim 1/p_{\rm dest}$.
 Therefore, in the case that the optical depth of the system, $\tau_0$, is much larger than $1/p_{\rm dest}$,
almost all Ly$\alpha$ photons are destroyed in the system.
Consequently, the force multiplier is independent of
$\tau_0$ but is capped at $1/p_{\rm dest}$.\footnote{{In a random walk caused by resonant scattering, the distance is proportional to the number of scatterings, as noted by \citealt{Dijkstra_review2019}. 
Photons escape the medium more easily during resonant scattering because the cross-section decreases as the photon frequency shifts away from the line center with an increasing number of scatterings.
}}

In contrast, for {$\tau_0 \ll 1/p_{\rm dest}$}, the destruction process is negligible
and thus the force multiplier can be  analytically obtained as $M_F^{\rm ana}(\tau_0)$. 
Therefore, the relationship between $\tau_0$ and $M_F$ in the presence of the destruction process is given by
\begin{align}
    M_{\rm F}(\tau_0,n_{\rm H})=\min\qty(M_{\rm F}^{\rm ana} (\tau_0), M_{\rm F}^{\rm ana} \qty(1/p_{\rm dest}(n_{\rm H}))).
    \label{eq:MFana_annihilation}
\end{align}
We now provide an overview of the behavior of $p_{\rm dest}$ by introducing an approximation:
\begin{align}
    p_{\rm dest}\equiv \bar{p}_{\rm dest}f(n_{\rm H})\approx\bar{p}_{\rm dest}\left(\frac{n_{\rm H}}{n_{\rm H}+n_{\rm H,crit1}}+\frac{n_{\rm H}}{n_{\rm H, crit2}}\right) ,
    \label{eq:p_approx}
\end{align}
where $\bar{p}_{\rm dest} \equiv A_{\rm 2s1s}/(3A_{\rm2p1s})\approx4.39\times10^{-9}$ is the typical value of $p_{\rm dest}$. The critical values $n_{\rm H,crit1}$ and $n_{\rm H,crit2}$ are defined as $n_{\rm H,crit1}\equiv n_{\rm   H}A_{\rm 2s1s}/C_{\rm 2s2p}\approx 1.55 \times 10^{8} (x_{\rm e}/10^{-4})^{-1} \mathrm{~cm}^{-3}$ and $n_{\rm H,crit2}\equiv   A_{\rm 2s1s}/( (x_{\rm e}/10^{-4}) \gamma_{21}(\rm e)+\gamma_{21}(\rm H))/4\approx 1.75 \times 10^{12} /( (x_{\rm   e,}/10^{-4})+0.145)\mathrm{~cm}^{-3}$, respectively, determine the $n_{\rm HI}$-dependence of $p_{\rm dest}$.
In the limit of large $\tau_0$, Eq.\ref{eq:MFana_annihilation} can be rewritten as
\begin{equation}
    M_{\rm F}(\infty, n_{\rm H})\approx150f({n_{\rm H}})^{-1/3}T_{4}^{-1/6}. \label{eq:M_F_limit}
\end{equation}
{Static homogeneous models reproduce the limit in Fig.\ref{fig4}, where the force multiplier of each models (colored circles) converges to the value of Eq.\ref{eq:M_F_limit} (colored dot-dashed lines) in the limit of the large optical depth.}
The function $f(n_{\rm H})\equiv p_{\rm dest}/\bar{p}_{\rm dest}$ can be approximated as 
\begin{equation}
    f(n_{\rm H}) \approx
    \begin{cases}
      \frac{n_{\rm H}}{n_{\rm H, crit1}} & \text{if } n_{\rm H} < n_{\rm H, crit1} \\
      1 & \text{if } n_{\rm H, crit1}<n_{\rm H}<n_{\rm H, crit2} \\
      \frac{n_{\rm H}}{n_{\rm H, crit2}} & \text{if } n_{\rm H} > n_{\rm H, crit2}
    \end{cases}.
    \label{eq:approx_f}
\end{equation}
The approximation in Eq.\ref{eq:p_approx} with Eq.\ref{eq:approx_f} has an accuracy up to $10^{-2}$ in relative error.
Using Eq.\ref{eq:approx_f},
Eq.\ref{eq:M_F_limit} 
shows that in the number density range between $10^8\:{\rm cm^{-3}}$ and $10^{12}\:{\rm cm^{-3}}$, the terminal force multiplier is almost constant,
as long as the gas temperature does not change.
In other density ranges, 
it exhibits a weak dependency on $n_{\rm HI}$, proportional to $n_{\rm HI}^{1/3}$. 
This can explain the trend in the region with large $\tau_0$ shown in Fig.\ref{fig4}, where force multiplier depends on $n[{\rm cm^{-3}}]$ in the range of $10^6$ to $10^8$, but not in the range of $10^8$ to $10^{13}$.
Therefore, the force multiplier remains almost constant around 130 in our model.

In Section \ref{sec3.2}, we emphasized that the velocity gradient is less significant than the destruction process.
The reason for this is that Ly$\alpha$ photons penetrate only in a narrow region at the innermost region $(r-r_{\rm min}<r_{\rm min})$, where the velocity changes only slightly.
 The distance at which the velocity gradient becomes non-negligible is given by comparing the frequency shift induced by the velocity gradient with that caused by the thermal velocity dispersion,
$\Delta r_{\rm vg} = v_{\rm th} \left|dv/dr\right|^{-1}$.

The velocity distribution is considered as $v(r)\sim v_{\rm th} (r/ r_{\rm B})^{-1/2}$ in the innermost region, so we have
\begin{align}
    \Delta r_{\rm vg}
    \sim 3\times 10^{-5} r_{\rm B} 
    \label{eq:Dr_vg},
\end{align}
 where we use $r=10^{-3}r_{\rm B} (=r_{\rm min})$.
On the other hand, 
the typical distance, $\Delta r_{\rm pd}$,  that Ly$\alpha$ photons can penetrate corresponds to the length at which the optical depth of the line-center is $1/p_{\rm dest}$.
Thus, $\Delta r_{\rm pd}$ at the 
the innermost region is evaluated as
\begin{align}
    \Delta r_{\rm pd} \sim 2\times p_{\rm dest}^{-1} \tau_0^{-1} r_{\rm min} \simeq 4\times 10^5 \tau_0^{-1} r_{\rm B},
    \label{eq:Dr_pd}
\end{align}
{which is derived from comparing the optical depth from $r_{\rm min}$ to $r_{\rm min} + \Delta r_{\rm pd}$ with $\tau^* = 1/p_{\rm dest}$} where we assume $n_{\rm HI} \propto r^{-3/2}$ and
$p_{\rm dest}\sim 5.0 \times 10^{-9}$.
In our models,
$\Delta r_{\rm vg}\gg\Delta r_{\rm pd}$ holds
since  the optical depth, $\tau_0$, is
between $10^{10}$ and $10^{15}$.
Consequently, the velocity gradient effect can be negligible. 